\documentclass{revtex4}
\usepackage{graphicx}
\usepackage{fancyhdr}
\usepackage{amsmath}
\usepackage{color}

\begin{document}

\begin{flushright}
KEK-TH-1616
%{\tt hep-ph/10mmxxx}\\
\end{flushright}

\title{What can we learn from the 126 GeV Higgs boson  \\ for the Planck scale physics ?\\
- Hierarchy problem and the stability of the vacuum -
%\footnote{Talk presented at SCGT12 (Nagoya, Dec. 4-7, 2012) and at HPNP2013 (Toyama, Feb. 13-16, 2013)}
\footnote{Contribution to SCGT12 "KMI-GCOE Workshop on Strong Coupling Gauge
Theories in the LHC Perspective", 4-7 Dec. 2012, Nagoya University
and HPNP2013 "Toyama International Workshop on Higgs as a Probe of New Physics 2013", 13--16, Feb. 2013, Toyama University}
}


\author{Satoshi Iso}
\affiliation{Theory Center, 
Institute of Particles and Nuclear Studies \\
High Energy Accelerator Research Organization (KEK) 
\\1-1 Oho,
Tsukuba, Ibaraki, Japan 305-0801}

\begin{abstract}
The discovery of  the Higgs particle at around 126 GeV has given us a big hint towards the origin of the
Higgs potential. The running quartic self-coupling decreases  and crosses zero
somewhere in the very high energy scale. It is usually considered as a signal of the
instability of the standard model (SM) vacuum, but it can also indicate
a link between the physics in the electroweak scale and the Planck scale. 
Furthermore, the LHC experiments as well as the flavor physics experiments give strong constraints on the 
physics beyond the SM. It urges us to reconsider the widely taken approach to the physics
beyond the SM (BSM), namely the approach based on the gauge unification below the Planck scale and
the resulting hierarchy problem.
Motivated by the recent experiments, we first revisit the hierarchy problem
and consider an alternative appoach  based on a classical conformality
of the SM without the Higgs mass term.

In this talk, I review our recent proposal of 
a B-L  extension of the SM with a flat Higgs potential at the Planck scale \cite{IsoOrikasa,IOO}. 
This model can be an alternative solution to the hierarchy problem
as well as being phenomenologically viable to explain the neutrino oscillations and 
the baryon asymmetry of the universe.
With an assumption that the Higgs has a flat potential at the Planck scale,
we show that 
the B-L symmetry is radiatively broken at the TeV scale via the
Coleman-Weinberg mechanism, and it triggers the electroweak symmetry
breaking through a radiatively generated scalar mixing. The ratio of these
two breaking scales is dynamically determined by the B-L gauge coupling.
\end{abstract}

\maketitle

\thispagestyle{fancy}


%%%%%%%%%%%%%%%%%%%%%%%%%%%%%%%%%%
\section{Central dogma of particle physics}\label{aba:sec1}
In the LHC era, we  acquired various hints towards the physics beyond the SM.
The first hint is of course the mass of the recently discovered Higgs-like particle.
The value of 126 GeV is quite interesting because it is close to the boarder of the stability bound.
Given the vev of the Higgs at 246 GeV, its mass gives an information of the curvature of the potential
at the minimum. As the mass 126 GeV is smaller than 246 GeV, the Higgs potential is rather shallow
and unstable against the radiative corrections. 
The (in)stability of the SM vacuum can be investigated explicitly 
by looking at the running bahaviour of the 
quartic coupling. 
The beta function of the  quartic Higgs coupling $\lambda_H$ is given by
\begin{align}
\beta_H
%=\frac{d \lambda_H}{dt}
 = \frac{1}{16 \pi^2}
\left( 24 \lambda_H^2 - 6 Y_t^4 + \frac{9}{8}g^4 + 
\frac{3}{8}g_Y^4  \right).
\label{betaSM}
\end{align}
$Y_t$ is the top Yukawa coupling and $g, g_Y$ are $SU(2)_L, U(1)_Y$ gauge couplings. 
It is either positive or negative whether the negative contribution by the 
large top Yukawa coupling is compensated by the gauge couplings  and the Higgs quartic
coupling. The corresponding quartic coupling to the 126 GeV Higgs boson 
does not suffice to  compensate it and the beta function is negative. So the running quartic 
coupling crosses zero somewhere at the UV energy scale. 
It is very suggestive that 
the observed value of the Higgs mass is close to the stability bound up to the Planck scale \cite{Espinosa}
\begin{align}
M_h[\mbox{GeV}] > 129.2 + 1.8 \left( \frac{M_t[\mbox{GeV}]-173.2}{0.9}\right)
 -0.5  \left( \frac{\alpha_s(M_Z)-0.1184}{0.0007}\right) \pm 1.0_{th} .
\label{stability-bound}
\end{align}
When the Higgs mass is lighter than the above bound, new physcis must appear 
below the Planck scale. But if it lies just on the border of the stability bound, it
gives a big hint to the origin of the Higgs potential at the Planck scale\cite{FP1}. 

Another important information  is  that the LHC results are almost
consistent with the SM.  Furhtermore the precision experiments of the flavor physics, 
Babar, Belle and LHCb, gave stringent constraints on the physics beyond
the SM.  
Of course, in spite of the above rather unexpectedly good agreement with the SM, 
there exist phenomena which cannot be explained within the SM.
Nuetrino oscillation requires the dimension 5 operator $l \phi l\phi$ and a new
scale beyond the SM must be introduced. The baryon asymmetry of the universe
also requires an additional source of the CP violation. 
The SM anyway needs to be extended to explain these phenomana.

The most common appoach to go beyond the SM is based on a
unification of  the gauge couplings below the Planck scale, i.e. the GUT scale.
Then we need a natural explanation why the electroweak scale is much smaller than the
GUT scale. 
In order to solve the hierarchical structure of the scales, the supersymmetry is introduced.
Here I call the sequence of ideas from GUT to the hierarchy problem and the low energy supersymmetry
{\it the central dogma of particle physics}. 
In addition to solving the hierarchy problem, it can  improve the gauge coupling unification
as well as providing candidates of the dark matter particles.
But as the bonus we get or as the price we pay, it predicts many new particles at the TeV scale
and the recent experiments have  given strong constraints on the models
with low energy supersymmetry.

In such circumstances, it may be a good time to reconsider the central dogma of particle physics.
In this note, we take an  approach to the hierarchy problem suggested by Bardeen.
In the next section, we interpret the Bardeen's argument in terms of the renormalization group.
If we adopt the argument, the most natural mechanism to break the electreweak symmetry
is the Coleman-Weinberg (CW) mechanism. 
But we know that the CW mechanism does not work within the SM because of the large top
Yukawa coupling, so we need to extend the SM. In section 3, we introduce our model, a classically
conformal $B-L$ extension of the SM and then discuss the dynamics of the model.

\section{Bardeen's argument of the hierarchy problem} 
We pay a special attention to the almost scale invariance of the SM.
At the classical level, the SM Lagrangian is conformal
invariant except for the Higgs mass term.
Bardeen  argued \cite{Bardeen} that once the classical conformal invariance and its minimal violation
by quantum anomalies are imposed on the SM, it may be free from the quadratic divergences.

Bardeen's argument on the hierarchy problem may be interpreted as follows \cite{AokiIso}.
We classify divergences of the scalar mass term in the SM into the following 3 classes, 
\begin{itemize}
\item quadratic divergences: $\Lambda^2$
\item logarithmic divergences with a small coefficient: $m^2 \log (\Lambda/\mu)$
\item logarithmic divergences with a large coefficient: $M^2 \log (\Lambda/\mu)$
\end{itemize}
The logarithmic divergences are operative both in the UV and the IR. In particular, it  controls a
running of coupling constants and is observable. On the other hand, the quadratic divergence can be
always removed by a subtraction. Once subtracted, it no longer appears in observable quantities.
In this sense, it gives a boundary condition of a quantity in the IR theory at the UV energy scale
where the IR theory is connected with a UV completion theory. 
Indeed, the RGE of a Higgs mass term $m^2$  in the SM 
\begin{align}
V(H)= - m^2 H^\dagger H + \lambda_H (H^\dagger H )^2
\end{align}
is approximately given by
\begin{align}
\frac{dm^2}{dt}=\frac{m^2}{16 \pi^2}
\left( 12 \lambda_H + 6 Y_t^2 - \frac{9}{2}g^2
-\frac{3}{2}g_Y^2 
\right).
\end{align}
The quadratic divergence is subtracted by
 a boundary condition either at the IR or UV scale. Once 
the initial condition of the RGE is given at the UV scale, it is no longer operative in the IR.
The RGE shows that the mass term $m^2$ is multiplicatively renormalized.
If it is zero at a UV scale $M_{UV}$, it continues to be zero at lower energy scales.
In this sense, the quadratic divergence is not the issue 
in the IR effective theory, but the issue in the UV completion theory. 
Hence if the SM (and its extension at the TeV scale)
 is directly connected with a UV completion theory at the Planck scale physics, 
the hierarchy problem turns out to be a problem of the boundary condition at the UV
scale. 
If the UV completion theory is an ordinary field theory, it will be 
difficult to protect the masslessness of the scalar particle 
against radiative corrections by massive particles of the UV scale
unless we introduce, e.g.  the low-energy supersymmetry. 
But in the string theory, symmetry is sometimes enhanced on a moduli space and 
massless scalars can survive even without supersymmetry. 
Also discrete symmetry like T-duality, which is invisible in the low
energy effective theory, may prohibit a generation of  potential at the string scale.

The multiplicative renormalization of the Higgs mass term is violated by
 a mixing with a massive field in the loop. If the massive field aquires its mass
in a different mechanism with the EWSB,
the Higgs mass has a logarithmic divergence 
\begin{align}
\delta m^2 \sim  \frac{\lambda^2 M^2}{16\pi^2} \log(\Lambda^2/m^2)
\label{RGEm2}
\end{align}
which modifies  the RGE  as
\begin{align}
\frac{dm^2}{dt}=\frac{m^2}{16 \pi^2}
\left( 12 \lambda_H + 6 Y_t^2 - \frac{9}{2}g^2
-\frac{3}{2}g_Y^2 
\right)  + \frac{M^2}{8\pi^2} \lambda^2.
\label{RGEm}
\end{align}
The last term corresponds to the logarithmic divergence with a large coefficient. 
 The coefficient $M^2$ has nothing to do with the mass of the Higgs $m^2$, and
it violates the multiplicativity of the Higgs mass.
Thus the hierarchy problem, namely the stability of the EWSB scale, is 
caused by such a mixing of relevant operators (mass terms) with  hierarchical energy scales
$m \ll M$. In the Bardeen's argument, he also imposes an absence of intermediate scales
above the EW scale. 
The logarithmic divergence with a large coefficient (\ref{RGEm2}) is sometimes confused with
the quadratic divergence, but if the UV completion theory is something like a string theory,
they should be distinguished.

From the above considerations, 
the hierarchy problem  can be solved by imposing
the following two different conditions;
\begin{itemize}
\item Correct boundary condition  at the UV (Planck) scale $M_{pl}$
\item Absence of mixings in  intermediate scales below $M_{pl}$ 
\end{itemize}
The first condition subtracts the quadratic divergence at the Planck scale.
It must be solved in the UV completion theory such as the string theory. 
The most natural boundary condition is that scalar fields
which appear in the low energy physics are massless at the Planck scale.
On the other hand, the second condition assures the absence of 
logarithmic divergences with large coefficients. 
Even if the scalars are massless at the Planck scale, they receive large radiative 
corrections from the mixing with other relevant operators.
Without a cancellation mechanism like the supersymmetry,
we need to impose an absence of 
 intermediate scales  between EW (or TeV) and Planck scales. 
Hence all symmetries are broken either at the Planck scale or near the EW scale.
Especially, the breakings of the supersymmetry or the grand unification of gauge coupling
should occur at the Planck scale.
This second condition is also emphasized in the Bardeen's argument \cite{Bardeen}.
In such a scenario, Planck scale physics is directly connected with the electroweak physics
 \cite{shapo}.

Hence a natural boundary condition of the mass term 
at the UV cut-off scale, e.g. $M_{Pl}$, is
\begin{align}
 m^2(M_{Pl}) =0.
\label{boundary-condition}
\end{align}
This is the condition of the classical conformality of the BSM.
The condition (\ref{boundary-condition}) must be justified in the UV completion theory, and
from the low energy effective theory point of view, it is just imposed
as a boundary condition
\footnote{The condition (\ref{boundary-condition}) may look similar to the Veltman condtion
\cite{Veltman}, but
they are conceptually different at all. 
In the Veltman condition, the quadratic divergence is considered to be cancelled between
various contributions of bosons and fermions. Such a cancellation occurs
in a very special situation of the IR physics. On the contrary, the condition (\ref{boundary-condition})
is independent of the matter content in the IR, and robust against a change of scales.
}.

\section{Flat potential at the Planck scale}
The condition (\ref{boundary-condition}) restricts the form of the Higgs potential as
\begin{align}
V(H)=\lambda_H (H^\dagger H)^2.
\end{align}
Here $\lambda_H$ is the running coupling and the RG improved effecitve potential
is given by making the coupling $\lambda_H (H)$ field dependent. 
The mass term is not generated even in the IR as discussed in the previous section
once the boundary condtion (\ref{boundary-condition}) is imposed at the boundary with
the UV completion theory. The mass of the Higgs at 126 GeV suggests that the running coupling
becomes asymptotically vanishing near the Planck scale. The current bounds 
(\ref{stability-bound}) is a bit heavier than the experimental data, but in this note,
we assume that the Higgs quartic coupling vanishes at the Planck scale. Hence
\begin{align}
V(H)=0 \mbox{ at the Planck scale}.
\label{flatpotential}
\end{align}
The condition may connect the SM in the IR with the string theory 
in the UV.
Now we have to solve two problems.
The first is whether we can construct a phenomenologically viable model starting from the 
condtion of the flatness of the Higgs potential (\ref{flatpotential}), and the second
is to derive such a boundary condtion from the UV completion theory such as a string theory.
Supersymmetry or grand unification, if exists, are broken at the Planck scale.
In the following we focus on the first problem by proposing a B-L extension of the SM
with a flat potential at the Planck scale.
 The second issue is left for future investigations.

Since the IR theory is assumed to have the boundary condition (\ref{flatpotential}),
the electroweak symmetry breaking should occur radiatively, namely the Coleman-Weinberg 
mechanism. However, it is now well-known that the CW mechanism cannot occur within
the SM
because of the large top-Yukawa coupling. Indeed, the CW mechanism is realized 
only when the beta-function of the quartic scalar coupling is positive and the
running quartic coupling crosses zero somewhere in the IR.
But as we saw, the beta function of the quartic Higgs coupling
is positive in the SM and its behavior is opposite to the 
CW mechanism. 
Hence, in order to realize the EWSB, we need an additional sector in which the symmetry
is broken radiatively by the CW mechanism and whose symmetry breaking 
triggers the EWSB.
In the next section, we introduce our model, namely a B-L extension of the SM with a flat potential
at the Planck scale.

\section{B-L extension of the SM with flat potential at Planck}
The idea to utilize the CW mechanism to solve the hierarchy problem was first
modelled by Meissner and Nicolai \cite{MaNi} (see also \cite{Dias}).
In addition to the SM particles, they introduced
right-handed neutrinos and a SM singlet scalar $\Phi$. 
Inspired by the work, we proposed a minimal 
phenomenologically viable model \cite{IOO}. 
It is the minimal B-L model \cite{B-L}, but with a classical conformality.
The model is similar to the one proposed by Meissner and Nicolai \cite{MaNi}, but the difference
is whether the B-L symmetry is gauged or not. 
In a recent paper we further showed that by imposing the
flatness (\ref{flatpotential}) of the Higgs potential at the Planck scale
the B-L breaking scale is related with the EWSB scale.
The ratio of two scales is dynamically determined by the B-L gauge coupling
and the B-L breaking scale is naturally constrained to be around TeV scale \cite{B-L2}
for a not so small B-L gauge coupling.

Besides the SM particles the model consists of the B-L gauge field with the gauge
coupling $g_{B-L}$, right-handed nuetrinos $\nu_R^i$ ($i=1,2,3$ denotes the generation index) 
and a SM singlet complex scalar field $\Phi$ with two units of the B-L charge. 
The model is anomaly free. The Lagrangian contains Majorana Yukawa coupling 
 $\sim Y_N^i \Phi \bar{\nu}_R^{ic} \nu_R^i$,
and the see-saw mechanism gives masses to the left-handed neutrinos once the scalar $\Phi$ acquires vev.

\section{Symmetry breakings of B-L and EW}
Since the B-L gauge symmetry is broken by the CW mechanism, the breaking scale
is correlated with the quartic coupling $\lambda_\Phi$ at the UV scale. 
Its running is described by 
\begin{equation}
 \frac{d\lambda_\Phi}{dt}=\frac{1}{16\pi^2}\left(
  20\lambda_\Phi^2 -\frac{1}{2}Tr\left[Y_N^4\right]
  +96g_{B-L}^4+ \cdots  \right).
%\lambda_\Phi\left(2Tr\left[Y_N^2\right]  -48g_{B-L}^2\right)\right).
\label{RGElamphi}
\end{equation}
If the Mayorana Yukawa coupling is not so large, the beta function is positive.
The typical behavior of the running $\lambda_\phi$ is drawn in Fig. 1.
It crosses zero at a lower energy scale $M_0$, then the B-L symmetry is broken at
$M_{B-L} \sim M_0 \exp(-1/4)$ through the CW mechanism\cite{IsoOrikasa}. 

%%%%%%%%%%%%%%%%%%%%%%%%%%%%%%%%%%%%%%%%%%%%%%%%
\begin{figure}[t]\begin{center}
\includegraphics[scale=.6]{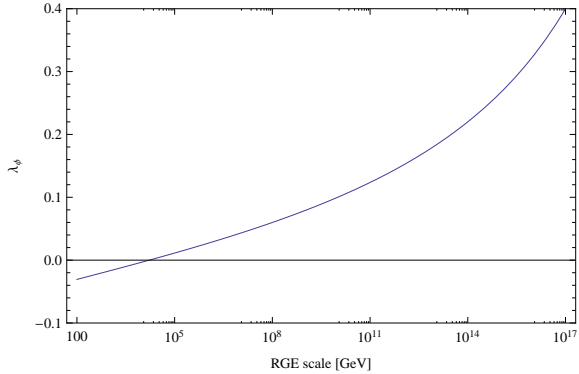}
\label{fig:self}
\caption{
RG evolution of the self-coupling $\lambda_\phi$ of a SM singlet scalar $\phi$.
Since the $\beta$ function is positive, the running coupling crosses zero
at a lower energy scale.
}
\end{center}
\end{figure}
%%%%%%%%%%%%%%%%%%%%%%%%%%%%%%%%%%%%%%%%%%%%%%%%
As shown in the paper \cite{IOO}, 
the ratio of the scalar boson mass to the B-L gauge boson mass is given  by
\begin{align}
\left( \frac{m_\Phi}{m_{Z'}}   \right)^2 \sim \frac{6}{\pi}  \alpha_{B-L} \ll 1.  % \mbox{ at } {M_{B-L}}. 
\label{massrelation}
\end{align}
The condition that the B-L gauge coupling does not diverge  up to the Planck scale
requires $\alpha_{B-L} < 0.015$ at $M_{B-L}$. Hence
the scalar boson becomes  lighter than the B-L gauge boson, $m^2_\Phi < 0.03 \ m^2_{Z'}$.
Such a very light scalar boson  is a general prediction of the CW mechanism.

The EWSB is triggered by the B-L breaking.
The flatness condition (\ref{flatpotential}) of the Higgs potential predicts
an absence of the scalar mixing at the Planck scale. Hence B-L and EW sectors are
decoupled each other in the UV. But since the matter fields are coupled to both $U(1)_Y$ and $U(1)_{B-L}$,
these two sectors become mixed through the $U(1)$-mixing. As a result,
the scalar mixing term $\lambda_{mix} (H^\dagger H) (\Phi^\dagger \Phi)$ appears
in the IR. It is interesting that a very small negative mixing $\lambda_{mix}$ is always 
generated irrespective of the details of other parameters once
we assume the flatness condition $\lambda_{mix}(M_{pl})=0$.
%We can see that a very small negative mixing is radiatively induced in  IR scale.
%The smallness and negativeness can be easily understood as follows.
%The RGE of $\lambda_{mix}$ is given by
%\begin{align}
%\frac{d\lambda_{mix}}{dt} \sim \frac{3}{4 \pi^2} g_{mix}^2g_{B-L}^2 + \lambda_{mix} (\cdots).
%\label{RGElammix-sim}
%\end{align}
%%%%%%%%%%%%%%%%%%%%%%%%%%%%%%%%%%%%%%%%%%%%%%%%
%\begin{figure}[t]\begin{center}
%\includegraphics[scale=.6]{lambdamix.eps}
%\caption{
%RG evolution of scalar mixing between a SM singlet $\Phi$ and the Higgs $H$.
%Starting from zero mixing at the UV scale, a small negative mixing is radiatively
%generated at a lower energy scale. This mixing triggers the EWSB.
%}
%\end{center}
%\end{figure}
%%%%%%%%%%%%%%%%%%%%%%%%%%%%%%%%%%%%%%%%%%%%%%%%%
%
%If there were no gauge mixing between $U(1)_Y$ and $U(1)_{B-L}$  gauge fields,
%the scalar mixing term would  never be generated radiatively when we start with the 
%condition $\lambda_{mix}(M_{pl})=0$. But because of the first term, the beta function becomes
%positive as long as the second term is negligible to the first, and a negative scalar
%mixing is generated in the IR. The smallness is guaranteed since the first term 
%is a higher power of gauge couplings.
By solving the RGE \cite{IsoOrikasa} we showed that the scalar mixing term is dynamically generated
around $\lambda_{mix} \sim - 4 \times 10^{-4}$. 
If the $\Phi$ field acquires a VEV $\langle \Phi \rangle =M_{B-L}$, the mixing term $\lambda_{mix} (H^\dagger H)(\Phi^\dagger \Phi)$
gives an effective mass term of the Higgs field.
Since  the coefficient $\lambda_{mix}$ is negative, the EWSB is triggered and the Higgs VEV is given by 
\begin{align}
v= \langle H \rangle = \sqrt{\frac{|\lambda_{mix}|}{\lambda_H}}  M_{B-L} 
\label{Higgsvev}
\end{align}
This gives a ratio between the EWSB scale to the B-L symmetry breaking scale.
The scalar mixing is determined in terms of the gauge couplings, so the ratio
of two breaking scales is also determined dynacamilly in terms of the gauge coupling $g_{B-L}$.

\section{Model predictions}
The dyanamics of the model is controlled by two  parameters, $g_{B-L}$ and $\lambda_{\Phi}$,
which determines the two breaking scales of B-L and EW.
%%%%%%%%%%%%%%%%%%%%%%%%%%%%%%%%%%%%%%%%%%%%%%%%
\begin{figure}[t]
\label{figure3}
\begin{center}
\includegraphics[scale=.8]{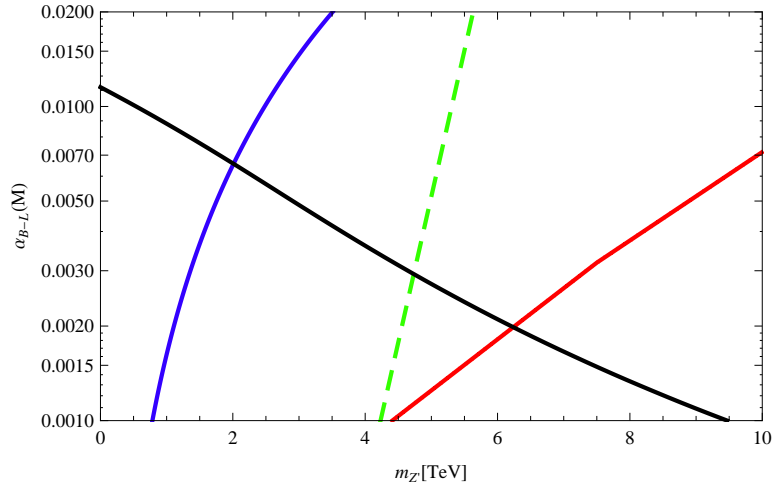}
\caption{Model prediction is drawn in the black line (from top left to down right).
The $B-L$ gauge coupling $\alpha_{B-L}$ and  the  gauge boson mass $m_{Z'}$
are related because of the  flat potential assumption at the Planck scale. 
The left side of the most left solid line in blue has been already excluded by 
 the LEP experiment.
The left of the dashed line can be explored in the 5-$\sigma$ significance 
 at the LHC with $\sqrt{s}$=14 TeV and an integrated 
 luminosity 100 fb$^{-1}$.
The left of the most right solid line (in red) can be explored at the ILC with 
 $\sqrt{s}$=1 TeV, assuming 1\% accuracy. 
}
\end{center}
\end{figure}
%%%%%%%%%%%%%%%%%%%%%%%%%%%%%%%%%%%%%%%%%%%%%%%%
The experimental input $v=246$ GeV gives a relation between these two
and the dynamics of the model is essentially described by a single parameter.
The figure 2 shows the prediction of our model.
The vertical axis is the strength of $\alpha_{B-L}$ and the horizontal axis is the
mass of the B-L gauge boson.
The black line (from top left to down right) shows the prediction of our model. 
If an extra $U(1)$ gauge boson and a SM singlet scalar are  found in the future,
the prediction of our model is the mass relation (\ref{massrelation}), e.g.,
$m_\phi \sim 0.1 \ m_{Z'}$
 for $\alpha_{B-L} \sim 0.005$.
The CW mechanism in the B-L sector predicts  a much lighter 
 SM singlet Higgs boson than the extra $U(1)$ gauge boson.
It is different from the ordinary TeV scale B-L model where the symmetry 
is broken by a negative squared mass term. 

Nuetrino oscillation is realized by the type I see-saw mechanism with small neutrino Yukawa couplings. 
Baryon number asymmetry of the universe may be generated through the TeV scale leptogenesis
with almost degenerate Majorana masses\cite{IOO3}. Furhter phenomenological issues such as
$U(1)$ mixing or the lepton number violation at the TeV scale are discussed in a separate paper.

%\begin{verbatim}

%% Create the reference section using BibTeX:
%%\bibliography{basename of .bib file}
%\begin{thebibliography}{99} % Use for 10-99 references
%\bibitem{DPF2009} http://www.dpf2009.wayne.edu/proceedings.php
%\bibitem{eConf} http://www.slac.stanford.edu/econf/
%\bibitem{templates-ref} http://www.slac.stanford.edu/econf/editors/eprint-template/instructions.html
%\bibitem{arXiv} http://arxiv.org/help
%
%\end{thebibliography}

\end{document}